\documentclass[preprint]{aastex}
\usepackage{graphicx,xfrac,amsmath,longtable,color}

\def\la{\mathrel{\hbox{\rlap{\hbox{\lower4pt\hbox{$\sim$}}}\hbox{$<$}}}}
\def\ga{\mathrel{\hbox{\rlap{\hbox{\lower4pt\hbox{$\sim$}}}\hbox{$>$}}}}

\def\kms{km~s$^{-1}$}

\def\dm15{{$\Delta$}$m_{15}$}

\def\v10{$V_{10}$(Si~II)}

\def\W575{$W(5750)$}
\def\W610{$W(6100)$}
\def\6100{the 6100~\AA\ absorption}

\def\msun{~M$_\odot$}

\def\m{~M$_\odot$}

\def\CaII7291{Ca {\sc II}] $\lambda\lambda$ 7291,7323\ }

\def\OI6300{[O {\sc I}] $\lambda\lambda$ 6300,6364\ }

\def\apj{Ap. J.}
\def\apjl{Ap. J. Lett.}
\def\apjs{Ap. J. Supp.}
\def\aj{Astron. J.}
\def\nat{Nature}
\def\mnras{Mon. Not. Royal. Soc.}
\def\aap{Astron. \& Astroph.}
\def\pasp{Pub. Astr. Soc. Pac.}
\def\apss{Ap. and Space. Sci.}

\def\nar{New Astronomy Reviews}
\def\actaa{Acta Astronomica}
\def\la{\mathrel{\hbox{\rlap{\hbox{\lower4pt\hbox{$\sim$}}}\hbox{$<$}}}}
\def\ga{\mathrel{\hbox{\rlap{\hbox{\lower4pt\hbox{$\sim$}}}\hbox{$>$}}}}

\def\kms{km~s$^{-1}$}

\def\dm15{{$\Delta$}$m_{15}$}

\def\v10{$V_{10}$(Si~II)}

\def\W575{$W(5750)$}
\def\W610{$W(6100)$}
\def\6100{the 6100~\AA\ absorption}

\def\msun{~M$_\odot$}

\def\m{~M$_\odot$}

\def\CaII7291{Ca {\sc II}] $\lambda\lambda$ 7291,7323\ }

\def\OI6300{[O {\sc I}] $\lambda\lambda$ 6300,6364\ }

\def\apj{Ap. J.}
\def\apjl{Ap. J. Lett.}
\def\apjs{Ap. J. Supp.}
\def\aj{Astron. J.}
\def\nat{Nature}
\def\mnras{Mon. Not. Royal. Soc.}
\def\aap{Astron. \& Astroph.}
\def\pasp{Pub. Astr. Soc. Pac.}
\def\apss{Ap. and Space. Sci.}

\shorttitle{Progenitors of Type Ia Supernovae}

\begin{document}

\title{White Dwarf/M Dwarf Binaries as Single Degenerate Progenitors
of Type Ia Supernovae}

\author{J. Craig Wheeler\altaffilmark{1}}
\authoremail{wheel@astro.as.utexas.edu}
\altaffiltext{1}{Department of Astronomy, University of Texas at Austin, 
Austin, TX, USA.}

\begin{abstract}

Limits on the companions of white dwarfs in the single 
degenerate scenario for the origin of Type Ia supernovae
(SN~Ia) have gotten increasingly tight, yet igniting a nearly
Chandrasekhar mass C/O white dwarf from a condition of
near hydrostatic equilibrium provides compelling agreement
with observed spectral evolution. The only type of 
non-degenerate stars that survive the tight limits,
$M_V \ga 8.4$ on the SNIa in SNR 0509-67.5 and $M_V \ga 9.5$
in the remnant of SN~1572 are M dwarfs. While M dwarfs are 
observed in cataclysmic variables, they have special properties 
that have not been considered in most work on the progenitors 
of SN~Ia: they have small but finite magnetic fields and they 
flare frequently. These properties are explored in the context 
of SNIa progenitors. White dwarf/M dwarf pairs may be sufficiently
plentiful to provide, in principle, an adequate rate of 
explosions even with slow orbital evolution due to magnetic
braking or gravitational radiation. Even modest magnetic
fields on the white dwarf and M dwarf will yield adequate
torques to lock the two stars together, resulting in a
slowly rotating white dwarf, with the magnetic poles
pointing at one another in the orbital plane. The mass 
loss will be channeled by a ``magnetic bottle" connecting
the two stars, landing on a concentrated polar area
on the white dwarf. This enhances the effective rate of 
accretion compared to spherical accretion. Luminosity
from accretion and hydrogen burning on the surface of 
the white dwarf may induce self-excited mass transfer. 
The combined effects of self-excited mass loss, polar
accretion, and magnetic inhibition of mixing of accretion
layers give possible means to beat the ``nova limit" 
and grow the white dwarf to the Chandrasekhar mass even 
at rather moderate mass accretion rates.

\end{abstract}



\section{Introduction}

There are two principal scenarios for producing Type Ia
supernovae (SN~Ia): in the single-degenerate (SD) model,
mass is transfered from a non-degenerate star onto a white 
dwarf (Wheeler \& Hansen 1971; Whelan \& Iben 1973); in the 
the double-degenerate (DD) model, two white dwarfs merge
(Tutikov \& Yungelson 1979; Iben \& Tutukov 1984; Webbink 1984). 

An important constraint
on any model is that it provide for the observed spectral
evolution. In the SD model, a white dwarf is grown slowly
to the point of carbon ignition and runaway at a central density
somewhat in excess of $10^9$ g cm$^{-3}$. Current models for the
ensuing thermonuclear explosion have a critical free parameter,
the density at which subsonic deflagration burning makes a 
transition to supersonic detonation. With that parametrization,
plausible models provide a very good first order agreement with 
the observed color and spectral evolution from the UV (H\"oflich, 
Wheeler \& Thielemann 1998) through the optical (H\"oflich et al. 
2006, 2011; Blondin et al. 2011) to the NIR (Wheeler et al. 1998) 
and from pre-maximum through the nebular phase (Kozma et al. 2005;
Gerardy et al. 2007) into the young supernova remnant phase 
(Badenes et al. 2006, 2007). In addition, important progress is being 
made on a physical understanding of the physics of the deflagration 
to detonation transition (Poludnenko \& Oran 2010, 2011; Poludnenko, 
Gardiner, \& Oran 2011). 

In contrast, the DD merger process is intrinsically messy and 
asymmetric. Early attempts to compute light curves and spectra 
from the models have not been especially successful (Fryer 
et al.  2010; Bloom et al. 2012). It is possible that the
disruption of the smaller mass white dwarf is slower than some 
models have predicted. Dan et al. (2011; see also Dan et al. 2012)
argue that the disruption may require tens of orbits. Whether 
this is sufficiently slow to yield carbon ignition anything like 
that invoked in the SD models, and hence a respectable spectral 
evolution, remains to be seen.  More effort will go into both 
classes of models, but at this writing, spectral evolution models 
based on secular evolution to degenerate ignition of a C/O white 
dwarf very near the Chandrasekhar mass are preferred, and that 
condition points to the SD model.

In the SD model, an associated constraint is the expectation that 
the mass transfer rate must be sufficiently rapid that accretion leads 
to non-degenerate shell burning on the surface of the white dwarf 
in order to avoid classical nova explosions that eject the 
accreted matter and, probably, some of the white dwarf material 
as well (Nomoto 1982; Iben 1982; Fujimoto 1982; Shen \& Bildsten 2008; 
and references therein). For solar abundances, this constraint 
requires the white dwarf to be bright and hot, qualities exhibited 
by the super-soft X-ray sources (SSS; van den Heuvel et al. 1992; 
Kahabka \& van den Heuvel 1997). While too few SSS are seen to 
account for either SD or DD progenitors (Di Stefano 2010a,b), 
Nielsen et al. (2012) and  Wheeler \& Pooley (2012) have argued 
that local circumstellar matter 
may easily block the emission of the predicted soft X-rays. Livio \& 
Truran (1992) summarized arguments that transfer of helium-rich matter 
might alleviate the problem of nova explosions and associated mass 
stripping from the white dwarf. Chatzopoulos, Robinson \& Wheeler 
(2012) have argued that that in close, tidally-locked binaries, 
the secondary stars are prone to rotationally-induced mixing and 
nearly homogeneous evolution, with associated enhancement of the 
helium at the expense of hydrogen. This may also promote evolution 
to explosion in environments that would otherwise seem unconducive.  
Another possibility is that the systematics of burning on the 
surface of the white dwarf are different than previously thought.
Starrfield et al. (2012) argue that nova explosions require
the mixing of white dwarf matter into the outer layers. In
the absence of that mixing, they argue that AGB-like thin shell 
burning instabilities produce unstable burning, but little mass 
loss for a wide variety of conditions. We will explore all these 
issues below.

Published models that satisfy the constraint that nova explosions 
should be avoided require the mass-transferring secondary star to 
be a moderately massive main sequence star, a sub-giant or giant 
star. Searches for any left-over secondary star in young remnants
of SN~Ia have so far proven ambivalent (Ruiz-Lapuente et al. 2004;
Gonz{\'a}lez Hern{\'a}ndez et al. 2009; Gonz{\'a}lez Hern{\'a}ndez 
et al. 2012; Kerzendorf et al. 2012a). 
The recent advent of SN 2011fe, an apparently normal 
``plain vanilla" SN~Ia, has provided new constraints on the progenitor 
systems of such supernovae. Nugent et al. (2011) argue that 
lack of light-curve contamination implies that the secondary 
star was not a red giant, and more likely to be a main sequence 
star. Li et al. (2011) use archival images to put limits on 
the companion and rule out luminous red giants and almost all 
helium-star models. Bloom et al. (2012) show that the exploding 
star was a white dwarf, as expected, and that the secondary 
star was likely to have had a radius less than 0.1 that of the 
Sun, excluding companion red-giant and solar-mass main-sequence 
stars that fill their Roche lobes.  

SNR 0509-67.5 in the LMC was established by the detection of 
scattered, time-delayed spectra to be a SN~Ia of the SN~1991T spectral
subclass that exploded about 400 years ago (Rest et al. 2008). 
Schaefer \& Pagnota (2012) examined deep HST images of this remnant
to put even tighter limits on the progenitor system of this explosion. 
They found that any surviving secondary star must be dimmer than $M_V \sim
8.4$ mag, ruling out basically all published SD models, including 
those with companion main sequence stars of greater than about 1\msun, 
sub-giants, giants, and those involving the stripped cores of evolved 
stars. While one might adopt the dodge that this was a single event 
responsible for a somewhat peculiar and ill-understood sub-class of 
SN~Ia (SN~1991T-like), and hence not typical of ``plain vanilla" 
SN~Ia, we assume here that this limit pertains to SD models of core 
normal events (Branch et al. (2006). Even tighter limits have been 
put on the remnant of SN~1572 with long HST exposures (Kerzendorf et al.
2012b; Schmidt private communication, 2012): 
$M_V > 9.5$ for any stars with space velocities 
in excess of $v \sim 100$ \kms. Gonz{\'a}lez Hern{\'a}ndez et al. (2012) 
have put a limit on any companion to the exploding white dwarf in the 
remnant of SN~1006 of $M_V \sim 4.9$. Either SD models must be rejected 
for these systems, or some means must be found to impeach the current 
set of SD models. 

Schaefer \& Pagnotta (2012) argue that a lower limit can be set on the 
mass of a main sequence companion of 1.16\msun. This constraint 
is derived from the requirement that the mass transfer be sufficiently 
rapid to avoid the nova fate and hence unstable. Rapid mass transfer 
in turn requires that the Roche lobe should shink more rapidly than 
the radius of the mass-losing star. This requires that the mass ratio 
must be $> 5/6$ so that for a primary of 1.4\msun\ the secondary must 
be $> 1.16$\msun. A successful SD model will also have to elude this limit. 

Yet another limit to consider is the amount of hydrogen swept
from a companion that will pollute the nebular spectra 
(Livne, Tuchman \& Wheeler 1992; Marietta, Burrows \& Fryxell 2000;
Pakmor et al. 2008; Kasen 2010; Pan, Ricker \& Taam 2012). Leonard 
(2007) gives an upper limit of 0.01 \m\ for the mass of solar abundance 
material in the nebular spectra of two SN~Ia. This constrains the 
liklihood that any companion is filling its Roche Lobe.

White dwarfs and red dwarfs are the most common stellar components 
of galaxies. Their pairing is often seen in cataclysmic variables. 
M dwarfs are, by definition, fainter than $M_V > 8.4$ and so meet the 
stringent limit set by SNR 0509-67.5, and the looser limits from 
SN~2011fe and runaway companion stars in SN~1006 (Gonz{\'a}lez 
Hern{\'a}ndez et al. 2012) and SN~1572 (Ruiz Lapuente et al. 2004). 
Many M dwarfs are also dimmer than the limits on a companion in
SN~1572 by Kerzendorf  et al. (2012b).

A binary system emerging from a common envelope phase with a white 
dwarf of $M > 1.0$\msun\ and an M dwarf companion of $M > 0.4$\msun\ 
would satisfy the criterion of a total mass exceeding the
Chandrasekhar mass. Taken at face value, the constraint given above
for the lower limit of main sequence companions rules out M dwarfs 
as the single degenerate component, but M dwarfs have special 
properties that might allow them to avoid this constraint. Virtually 
all of the current models of SD systems are based on one-dimensional, 
spherically-symmetric, non-rotating, non-magnetic accretion that is 
undoubtedly incorrect, at least in detail, and does not describe
the reality of M dwarfs. The issue is whether or not the special 
properties of M dwarfs as fully convective, magnetic, flaring 
stars provide a different set of conditions than has been previously 
considered that might make them a serious candidate for the 
non-degenerate companions in the progenitors of SN~Ia.   

This paper is organized as follows. Section \ref{rates} discusses the
quantity and mass distributions of white dwarfs and M dwarfs in the 
Galaxy, the liklihood that they are paired, their rates of coalescence 
and the implications for the associated delay-time distribution. 
Section \ref{properties} discusses the special properties of
M dwarfs as dim, long-lived, magnetic, flaring, nearly or 
fully-convective stars. Section \ref{bottle} outlines the possible
implications that both stars, white dwarf and M dwarf, have
magnetic fields so that the combined magnetic structure should be
considered as the stars interact and exchange matter. Section
\ref{transfer} discusses the possible systematics of mass transfer
and mass and angular momentum loss from the system with the
assumption that both stars are locked to rotate at the orbital
period. Section \ref{irradiation} examines the possibility that a
close white dwarf, M dwarf pair can be subject to self-excited
mass transfer that could maintain a high transfer rate and
suppress nova-like shell flashes. Section \ref{post} discusses
the possible conditions of the M dwarf after a companion
explodes as a supernova and \S\ref{concl} summarizes the 
possible advantage of white dwarf, M dwarf pairs as progenitors
of SN~Ia and outlines important outstanding issues. 

\section{Rates}
\label{rates}

About 70\% of the stars in the Galaxy are M dwarfs. The number of 
M dwarfs in the Galaxy is thus $\sim10^{12}$ (Bochanski et al. 2010;
Bochanski, Hawley \& West 2011). 
From the PTF M dwarf study, Law et al. (2012) estimate that about 
1 M dwarf in 1000 has a white dwarf companion (Law et al. 2012). 
This suggests that there are $\sim10^9$ white dwarf/M dwarf (WD/M) pairs 
in the Galaxy. There are $\sim10^{10}$ white dwarfs in the Galaxy, 
17\% in the thin disk, 34\% in the thick disk and 49\% in the halo 
(Napiwotzki, 2009). Thus, to zeroth order, 10\% of all white dwarfs 
have M dwarf companions. From PanSTAARS data, Tonry (private 
communication, 2012) estimates the fraction to be $\sim20\%$.

An important critical factor is the mass distribution of the 
white dwarfs. For the hot ($>$ 12,000 K) DA field white dwarfs,
Kepler et al. (2008; see also Ferrario et al. 2005) 
estimate that most of the white dwarfs peak at
0.572\msun, that $\sim 19\%$ have a mass of $\sim 0.8$\msun\
and that $\sim 9\%$ have a mass of $\sim 1.1$\msun. These masses 
may be underestimated by $\sim 10\%$ (Falcon et al. 2010). Lower 
metallicity stars tend to make more massive white dwarfs 
(Willson 2000). These numbers suggest that there may be $\sim 10^8$ 
white dwarfs with M dwarf companions that have mass $\sim 1.1$ 
and about twice that many WD/M pairs with white dwarf masses of 
$\sim 0.8$\msun. The former only need to accrete $\sim 0.3$\msun\ 
to reach the Chandrasekhar limit and the latter only $\sim 0.6$\msun. 
A caveat is that the more massive white dwarfs may be composed of 
O/Ne/Mg. Whatever catastrophic endpoint such white dwarfs reach, 
it will not correspond to observed SN~Ia. 

With these numbers, one can crudely estimate that if the WD/M pairs 
coalesce in 10 Gyr, the resulting rate would correspond to $\sim 0.01$ 
event per year, in the right ball park to correspond to the SN~Ia rate. 
There are interesting questions associated with issues of how WD/M 
pairs form in terms of initial binary configurations and subsequent 
common envelope evolution.  In the current work, these issues are 
side stepped by invoking the observed existence of WD/M pairs. The 
rate of coalescence will depend on the distribution of orbital 
separations of the white dwarfs and M dwarfs, but this can also be 
determined empirically. Forthcoming data from PTF and PanSTAARS are 
greatly anticipated. 

\begin{figure}[htp]
\centering
\includegraphics[totalheight=0.5\textheight]{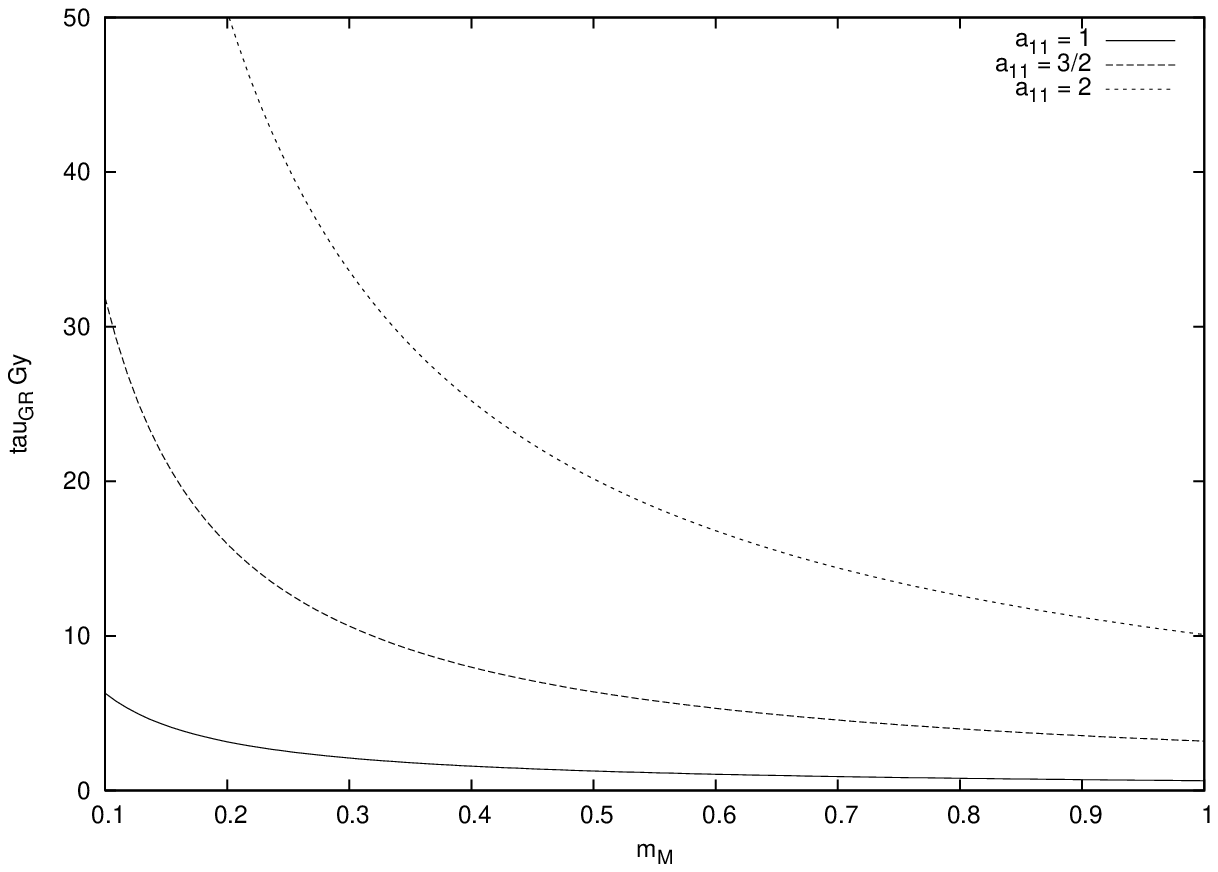}
\figcaption[taugr.ps]
{The time to inspiral, $\tau_{GR}$, in Gyr from a given initial 
separation, $a$, is given as a function of the M dwarf mass, $m_M$, 
for various values of $a$, assuming that the mass of the white dwarf 
is $M_{WD} = 1.4$\m. These curves represent upper limits for the
white dwarf and M dwarf to interact because winds from the M dwarf 
and from the ``magnetic bottle" (\S \ref{bottle}) and associated
loss of orbital angular momentum have been neglected.
\label{gr}}
\end{figure}

In this work, the issue of initial orbital separations will be 
treated in a qualitative manner. The timescale to reach contact 
will depend on the rate of dissipation of orbital angular 
momentum. Before interaction between the white dwarf and the 
M dwarf this will be driven by gravitational radiation or magnetic 
braking. Studies over the years have suggested that the latter 
is rather weaker than first models (Verbunt \& Zwaan 1981) suggested 
(Ivanova \& Taam 2003; Andronov, Pinsonneault \& Sills 2003; Willems et al.
2007; Chatzopoulos, Robinson \& Wheeler 2012), especially for systems
for which both stars are magnetic, as will be invoked here (Li, Wu 
\& Wickramasinghe 1994b). For simplicity, only gravitational radiation 
loss will be adopted here. The time to reach contact or coalesce from 
a given initial separation is given by Paczy{\'n}ski (1971) as:
\begin{equation}
\label{tau-gr}
\tau_{GR} = 0.63 {\rm Gy} \frac{a_{11}^4}{m_M m_{WD} m_{TOT}}, 
\end{equation}
where the $m_M$ is the mass of the M star, $m_{WD}$ the mass of 
the white dwarf, and $m_{TOT}$ the total mass, and we adopt the 
convention that masses expressed in units of the solar mass are 
denoted by lower case $m$. Here and throughout, we adopt the 
convention that subscript $M$ denotes the M dwarf and subscript 
$WD$ denotes the white dwarf. Figure \ref{gr} gives $\tau_{GR}$ 
as a function of $m_M$ far various assumed values of a. To coalesce 
in less than 10 Gy with gravitational radiation the only mechanism
of dissipation of orbital angular momentum requires 
$a_{11} \la 2(m_M m_{WD} m_{TOT})^{1/4}$, corresponding to a period 
of less than $P \sim 13~{\rm hr}~(m_M m_{WD})^{3/8}/m_{TOT}^{1/8}$. 
These are lower limits since magnetic braking due to winds from
the M dwarf and possible loss of angular momentum associated with the
magnetic interaction of the two stars (\S \ref{transfer}) have
been neglected. It remains to be seen whether or not a sufficient 
number of WD/M pairs are produced with sufficiently small orbits 
to give a high rate of coalescence. The statistics of the cataclysmic 
variable orbital period distribution tend to be rather thin for 
larger orbital separations (Woudt et al. 2012). 

Another constraint on a model for SN~Ia is the delay time distribution
(DTD) that falls off roughly like $t^{-1}$ (Maoz \& Manucci 2012). 
In most models for the DTD in the SD scenario, the DTD drops rather 
sharply after a few Gyr (Greggio 2005; Maoz \& Mannucci 2012). 
The reason is that for longer times and lower primary masses 
the primary produces smaller mass white dwarfs. The companion
thus needs to provide ever more mass to grow the white dwarf
to the Chandrasekar mass. In the standard SD model, secondary 
donors with low mass cannot transfer enough matter at sufficiently 
high rates to produce a SN~Ia. After a few Gyr, when the 
secondaries are not massive enough, the presumption is made
that SN~Ia cannot be produced, so the DTD declines. The goal
of the current paper is to explore whether or not that last 
assumption is correct. 

In the case where the secondary is an M star, then the 
DTD will be dictated by systematics rather similar to those for the 
DD model. The evolution time of the secondary is so long as to be 
irrelevant. The rate of coalescence will be determined by 
gravitational radiation (or perhaps magnetic braking) that drives 
the secondary into the condition of mass transfer (at or near 
filling its Roche lobe). From Equation \ref{tau-gr} the inspiral time 
is proportional to $a^4$. If the distribution of separations after
common envelope evolution is a power law, $n \propto a^\beta$, 
then the DTD will be $\propto t^{(\beta -3)/4}$, and if the
separation distribution is nearly flat in the log of the separation, 
$\beta=-1$, the DTD would be $dN/dt \propto t^{-1}$, as observed. 
Pritchet et al. (2008) describe a possibility when the time between 
the formation of the white dwarf and its explosion is short compared 
to the formation time of the white dwarf. In that case, the DTD is 
just proportional to the rate of formation of the white dwarfs. The 
rate of production limits the rate at which the inspiral can occur 
and the DTD may have a more shallow slope than would be the case if 
the gravitational radiation were the controlling factor. Following 
Pritchet et al. (2008) and Maoz \& Manucci (2012), one assumes 
$t \propto m^\delta$ from stellar evolution calculations and a power-law 
IMF $dN/dm \propto m^\lambda$. The DTD would then be $dN/dm \propto 
t^{(1+\lambda-\delta)/\delta}$. For $\delta \sim - 2.5$ and $\lambda 
\sim -2.35$ (Salpeter), the power law is about -1/2. This would 
represent a relatively slow fall off for short delay times, tending to 
the -1 slope for older systems. For M dwarfs, however, the IMF is much 
more shallow (next section). If $|\lambda|$ is small compared to 
$|1-\delta|$, then the slope would be about -1.4, not much different 
than the observed slope, within the uncetainties. Pritchet et al. argue 
that for the SD scenario to work about 1\% of all white dwarfs must 
be converted to SN~Ia, independent of mass. The ubiquity of WD/M pairs,
$\sim 10\%$ of all white dwarfs, raises the possibility that this 
may be true for this population. There are many issues here that can 
only be addressed with a proper binary synthesis calculation involving 
the common envelope and subsequent evolution of modest mass stars with 
M dwarf companions and associated observations of WD/M pairs at a 
variety of separations.

\section{Properties of M Dwarfs}
\label{properties}

The mass of M dwarfs as a function of spectral type
is given by Delfosse et al. (2000); M0V stars have
a mass of $\sim 0.6$\msun\ and those of M4V have
a mass of $\sim 0.2$\msun. Theoretical models have 
been used to estimate that the transition from 
radiative core to the fully convective limit occurs
at $\sim 0.35$ (Chabrier \& Baraffe 2000). This limit 
is based on one-dimensional, non-rotating, non-magnetic 
models and is probably incorrect at some level, but is 
useful for perspective. The structure of M dwarfs
may be affected by their presence in tidally-locked
binary systems (Chabrier, Gallardo \& Baraffe 2007).
The mass distribution of M dwarfs, the spatial density
versus mass, is essentially flat (Bochanski et al. 2010),
and will be assumed so in the following.

A special property of the M dwarfs is that they flare frequently
with associated coronal mass ejection (CME). The CMEs result 
in the loss of mass and represents a minimum rate for that process. 
The flare rate, $\nu_f$, depends on the energy of the flares. Hilton 
(2011; his figure 4.12) gives an opticaal flare rate of $\nu_f \sim 
0.1 E_{30}^{-0.7}$ hr$^{-1}$, where $E_{30}$ is the flare energy 
in units of $10^{30}$ erg. The mass in a given CME, $M_{CME}$, also 
scales with the flare energy. Aarnio et al. (2011; also Stassun, 
private communication, 2012) give the CME mass as $M_{CME} \sim 
10^{16} E_{30}^{0.7}$ g. The associated mass loss rate is then 
$\nu_f M_{CME} \sim 3\times10^{11}$ g s$^{-1}$, essentially independent 
of the flare energy. This mass loss rate corresponds to 
$\sim 5\times10^{-15}$\msun\ y$^{-1}$, close to that estimated by 
Cohen (2011) by other means. A more accurate estimate of the
energy slopes gives $\sim 10^{-13}--10^{-12}$\msun\ y$^{-1}$
(Aarnio, private communication, 2012). The flaring, CME  mass loss 
is thus rather small unless the presence of the white dwarf has 
some substantial effect. 

An intrinsically important aspect of M dwarfs, closely related to
their flare properties, is that they are magnetic (Donati et al.
2008; Stassun et al. 2010). There is a transition in field strength 
around spectral type MV4, corresponding to a mass of $\sim0.2$\msun
(Stassun et al. 2010; their Figure 7). For earlier spectral types 
and larger masses, the typical field strength is $\sim 100$ G. For 
spectral types of $\sim$ MV4 and later (and depending on the 
orbital period and other factors) the typical field strength is 
$\sim$ 1000 G. It is not at all clear that this variation in 
field strength with mass could be associated with mass loss 
during binary evolution, but that is an interesting possibility. 
It may also be true that the state of rotation, action of dynamos, 
even the transition from partially radiative to fully convective, 
may be altered for M dwarfs in close binary systems. 

\section{Magnetic White Dwarf, M Dwarf pairs: Magnetic Bottle}
\label{bottle}

While not all white dwarfs have measurably large magnetic
fields, many may have modest magnetic fields. Liebert,
Bergeron \& Holbert (2003) argue that over 10\% of white
dwarfs have fields in excess of $2\times10^6$ G. Here,
we will assume that a significant fraction of the white
dwarfs with M dwarf companions have fields on the order
of 1 MG. 

The WD/M binaries we envision are thus
related to intermediate polars or polars, but the 
important feature that the companion to the white dwarf is 
also magnetic. The fiducial field of the M dwarf will
be taken to be 100 G, but with the proviso noted above
that for later spectral types 1000 G may be more typical. 
For dipole fields and for conditions where the M star
nearly fills its Roche lobe, as will be discussed below, 
the field strength of the M star dipole field will be much 
stronger in the vicinity of the M dwarf than the 
field of the white dwarf field at the M star. The implication
is that it is important to consider the merged magnetic
field structure of the two stars.  

For two stars, each with substantial dipole field structures, the 
stable orientation is for the dipoles to be aligned, pointing
at one another in the orbital plane (King, Whitehurst \& Frank
1990; see also the extensive discussion of this and related
issues in Cambell 1983, 2010 and references therein and Wu \& 
Wickramashighe 1993; Li, Wu \& Wickramasinghe 1994a,b)). The torque 
between the two dipoles separated by a distance, a, is:
\begin{equation}
\label{torque}
\tau = \frac{\mu_{WD}\mu_M}{a^3} = \frac{B_{WD}R_{WD}^3 B_M R_M^3}{a^3},
\end{equation} 
where $\mu$ represents the magnetic moment, $B$ the surface magnetic 
field and $R$ the radius of each star. A heuristic estimate of the 
timescale to reach the stable point with the two dipoles pointing 
at one another is:
\begin{equation}
\label{time}
t_{mag} \sim \sqrt{\frac{I}{\tau}} = \sqrt{\frac{kMR^2}{\tau}},
\end{equation} 
where $I = kMR^2$ is the relevant moment of inertia and $k$ is 
the radius of gyration.
In practice, the two dipoles are likely to oscillate around the
equilibrium point. King, Whitehurst \& Frank (1990) give this
oscillation frequency as:
\begin{equation}
\omega \propto \left(\frac{\mu_{WD}\mu_{M}}{Ia^3}\right)^{1/2},
\end{equation}
where the coefficient of proportionality is a function of the angles
of the dipoles and I is that appropriate for the respective
stars. King et al. identify two modes of oscillation, a fast mode
and a slow mode, corresponding to the white dwarf and the secondary,
respectively, each with a somewhat different angular dependence. 
The oscillation amplitude of the white dwarf dipole tends to be
larger than for the lower mass secondary. In the current context,
$t_{mag} \sim 2\pi/\omega$ and we will take $t_{mag}$ to represent,
at least crudely, the damping time for the two dipoles to come
into near alignment. In the current work, the oscillation of
the dipoles will be ignored, but this is clearly a factor that
must be taken into account in any more detailed model. This
aspect will be pointed out where relevant below (\S \ref{irradiation}).

The torque between the two stars will increase as the two stars
approach one another. To illustrate the characteristic time
scales, we consider the condition that the M dwarf fills its
Roche lobe. As we will explore below, this may not necessarily
be the case. The M dwarf may also not be in the same thermal
equilibrium state that a star of that mass would be in 
isolation, but that does not affect the immediate estimates here.
For the separation corresponding to the M dwarf filling its
Roche lobe, $a_{RL}$, we adopt the simple expression given by 
Paczy{\'n}ski (1971) for the case when the secondary is substantially 
less massive than the white dwarf primary:
\begin{equation}
\label{Roche}
\frac{R_M}{a_{RL}} \sim 0.46 \left(\frac{M_M}{M_{TOT}}\right)^{1/3}.   
\end{equation}
Equation \ref{torque} then becomes:
\begin{equation}
\tau \sim 9.7\times10^{33}~B_{WD,6}B_{M,2}R_{WD,9}^3
    \left(\frac{M_M}{M_{TOT}}\right) {\rm dyne~cm},
\end{equation}
where $B_{WD,6}$ is the white dwarf surface dipole magnetic field
in units of $10^6$G, $B_{M,2}$ is the M dwarf surface dipole
field in units of 100 G, and $R_{WD,9}$ is the radius of the
white dwarf in units of $10^9$ cm. Note that in this approximation,
the torque does not depend on the radius of the M star. The 
characteristic relaxation timescale for the white dwarf is thus:
\begin{equation}
\label{time,wd}
t_{mag,WD} \sim 15~{\rm yr} \left(\frac{k_{WD}m_{TOT}}
   {B_{WD,6}B_{M,2}R_{WD,9}}\right)^{1/2}
     \left(\frac{M_{WD}}{M_M}\right)^{1/2}.    
\end{equation} 
The characteristic relaxation timescale for the M dwarf is:
\begin{equation}
\label{time,m}
t_{mag,M} \sim 1.5\times10^2~{\rm yr} \left(\frac{k_{M}m_{TOT}}
   {B_{WD,6}B_{M,2}R_{WD,9}}\right)^{1/2}
     \left(\frac{R_{M,10}}{R_{WD,9}}\right), 
\end{equation}
where $R_{M,10}$ is the radius of the M dwarf in units of $10^{10}$ cm.

While these estimates are only accurate to of order unity,
it is clear that the timescales for the dipoles to align
are quite short when the system is close to contact. The M dwarf
is expected to be brought into synchronous rotation by tides
(Zahn 1977, 1989). Absent any disk torque, as will be discussed below, 
the magnetic torque acting on the white dwarf will also bring
it into synchronism on roughly the timescale given by Eqn. 
\ref{time,wd} (Campbell 1983, 2010). One important implication 
is that the white dwarf may be in quite slow rotation, with a 
period of hours, in contrast with much thinking about SN~Ia for 
which the expectation is that the white dwarf is likely to be spun 
up to rather rapid rotation by accretion torques acting alone.

\begin{figure}[htp]
\centering
\includegraphics[totalheight=0.5\textheight]{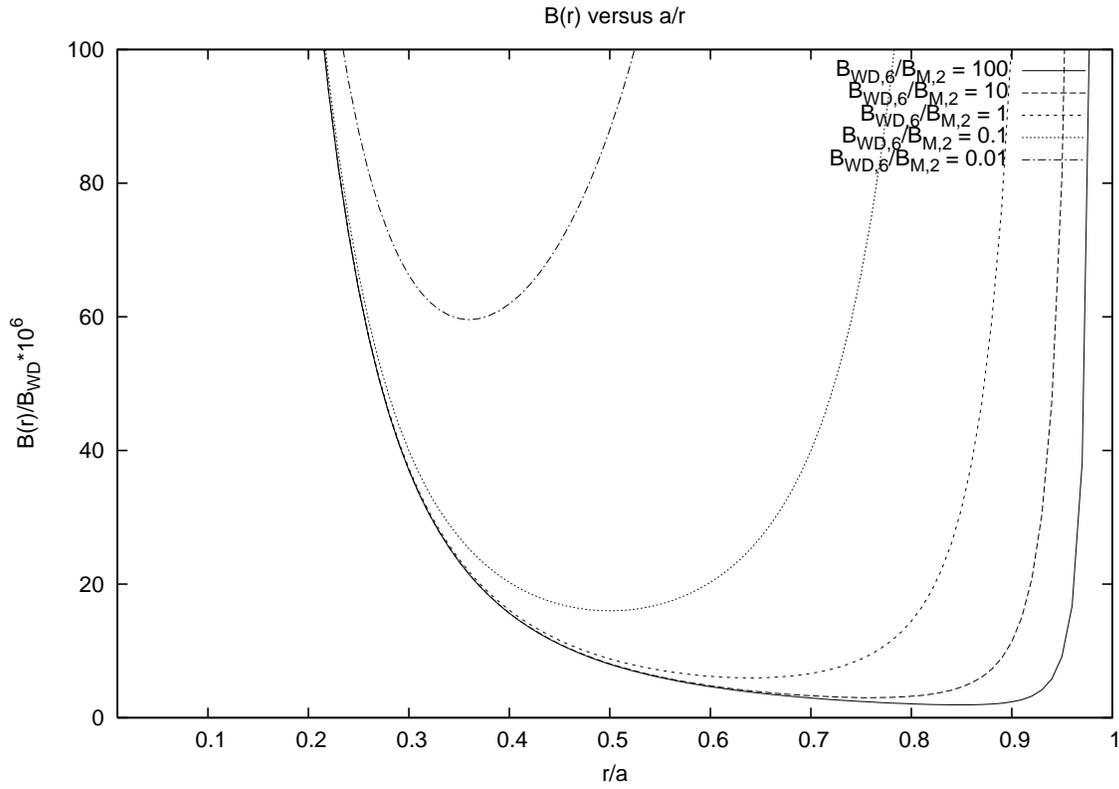}
\figcaption[dipolefield.ps]
{The ratio $B(r)/B_{WD}$ for two aligned dipoles a distance, a, apart
is given as a function of $r/a$ for various values of the ratio 
$B_{WD,6}/B_{M,2}$, where r is the distance from the center of the
white dwarf.
\label{dipoles}}
\end{figure}

Adopting the simplifying assumption that the two dipoles
align with one another, the field between the dipoles, along
the line of connection, can be described as:
\begin{equation}
B(r) = B_{WD}\frac{R_{WD}^3}{r^3} + B_M\frac{R_M^3}{(a - r)^3},
\end{equation}
where $a$ is the separation and $r$ is the distance from the white 
dwarf (by arbitrary choice). The minimum in the field will then be
at a distance from the white dwarf given by:
\begin{equation}
\label{min}
r = a \left(\frac{X - X^{1/2}}{X -1}\right),
\end{equation}
where the ratio of the magnetic moments is: 
\begin{equation}
X = \frac{B_{WD}R_{WD}^3}{B_M R_M^3} = 
    10 \frac{B_{WD,6}R_{WD,9}^3}{B_{M,2} R_{M,10}^3}.
\end{equation}
Figure \ref{dipoles} gives $B(r)/B_{WD}$ as a function of
$r/a$ for various values of the ratio $B_{WD}/B_M$. For 
very large $X$, $r \sim a$, and the minimum will be 
near the M dwarf. For small $X$, $r \sim X^{1/2} a$, and
the minimum will be near the white dwarf. Note that
the value of $X$ is sensitive to parameters and that
care must be taken for $X \sim 1$ in Equation \ref{min}, 
for which $r \sim 1/2 a$. 

The qualitative conclusion is that the combined magnetic
field will constitute a ``magnetic bottle" connecting the
two stars that will, among other things, channel mass
from the M dwarf to the white dwarf. In the simple picture
that the two stars orbit in synchronism with the orbit
and that the magnetic dipoles are aligned, this channel
will lead directly from the magnetic pole of the M star
to the magnetic pole of white dwarf. In practice, there will
be Coriolis forces and other effects that affect the flow
that we ignore for now. A more detailed examination of 
the mutual field distribution of oppositely aligned dipoles
normal to the orbital plane is given by Li, Wu \& Wickramasinghe
(1994a,b). They define ``dead zones" where field lines
are not free, but connect the two stars directly. In essence,
the ``magnetic bottle" is a large dead zone created by the
dipoles aligned in the orbital plane.


The existence of this ``magnetic bottle" configuration
has implications for the behavior of mass loss from
the system. Mass that might have been lost from the
M star in a wind or flares, will tend, instead, to
be captured and directed at the white dwarf. A model
under wide consideration is that proposed by Hachisu,
Kato \& Nomoto (1996; 2010) in which excess mass transferred
to the white dwarf is expelled in a wind. The critical
mass transfer rate to initiate such a wind is
$\dot{M} \sim 10^{-6}$ \msun\ y$^{-1}$. This wind is expected
to have a speed $\sim 1000$ \kms. The ram pressure
from such a wind on the scale of the binary system can be written:
\begin{equation}
P_{wind} \sim 2.5\times10^4 \frac{\dot{M}_{-6}v_8}{r_{11}^2} 
    {\rm dynes~cm^{-2}}, 
\end{equation}
where the mass loss rate is scaled to $\dot{M} = 10^{-6}$ 
\msun~y$^{-1}$, the velocity to 1000 \kms\ and 
the typical orbital size, $r$, to $10^{11}$ cm. The wind pressure
near the surface of the white dwarf will be $\sim10^4$ times
larger. The magnetic pressure due to the field of the white dwarf 
can be written:
\begin{equation}
P_{WD} \sim 4\times10^{10} B_{WD,6}^2 \left(\frac{R_{WD}}{r}\right)^{6} 
     {\rm dynes~cm^{-2}}. 
\end{equation}
The ratio of the magnetic to wind pressure near the surface of the
white dwarf can then be written:
\begin{equation}
\frac{P_{WD}}{P_{wind}} \sim 160 \frac{B_{WD,6}^2 R_{WD,9}^6}
     {\dot{M}_{-6}v_8}{r_{9}^{-4}} 
\end{equation}
The magnetic pressure will dominate over the wind pressure for
\begin{equation}
r_{9} \la 3.6 B_{WD,6}^{1/2}R_{WD,9}^{3/2}\dot{M}_{-6}^{-1/4}v_8^{-1/4}.
\end{equation}
The magnetic field may thus squelch, or at least substantially
interfere with, the wind out to several times the size of the white
dwarf. 

Whether the wind can then pick up again will depend, in 
part, on the presence of the M dwarf. If we assume that the 
M dwarf fills its Roche lobe in accord with Equation \ref{Roche},
the magnetic pressure due to the field of the M dwarf can be written:
\begin{equation}
P_M \sim 400 B_{M,2}^2 \left(\frac{R_M}{r}\right)^{6} {\rm dynes~cm^{-2}} 
  \sim 4 B_{M,2}^2 \left(\frac{M_M}{M_{TOT}}\right)^2 
     \left(\frac{a_{RL}}{r}\right)^{6} {\rm dynes~cm^{-2}}. 
\end{equation}
The ratio of the magnetic pressure from the M dwarf to that
of the (unimpeded) wind can then be written as:
\begin{equation}
\frac{P_{M}}{P_{wind}} \sim 1.6\times10^{-4}B_{M,2}^2 
   \left(\frac{M_M}{M_{TOT}}\right)^2 \dot{M}_{-6}^{-1}v_8^{-1}  
     \left(\frac{a_{RL}}{r}\right)^{6} r_{11}. 
\end{equation}
If we assume that $r \sim a_{RL}$ and that the radius of the M dwarf 
is proportional to its mass in Equation \ref{Roche} (see Equation 
\ref{Roche_separation}), then this ratio can be written:
and 
\begin{equation}
\frac{P_{M}}{P_{wind}} \sim 3.5\times10^{-4}B_{M,2}^2
   m_M^{10/3}m_{TOT}^{-4/3}\dot{M}_{-6}^{-1}v_8^{-1}.
\end{equation} 
The M dwarf alone would thus not be a substantial influence on
the hypothesized white dwarf wind unless the mass loss rate and
wind velocity were smaller than assumed. This might, however,
be the case, given that the magnetic field of the white dwarf may
itself interfere with the wind. The M dwarf would exert a
more significant effect on the white dwarf wind if its
field were bigger. This raises the issue of whether or not
the tendency for M dwarfs later than MV4 to have larger
magnetic fields might be manifested in a binary system
with mass loss from the M dwarf, not just as a statistical
property of the sample of field M dwarfs as discussed in
\S \ref{properties}. If the M dwarf field were larger by
a factor of 10, and the mass loss rate and wind velocity
were less by factors of 10 each, the M dwarf alone might
impede the flow of any wind from the white dwarf. 

The presence of the white dwarf and the M dwarf with its magnetic
field may thus bottle up any wind from the white dwarf. Note that 
the slow rotation of the white dwarf will also tend to prevent 
any magnetic ``propeller" effects on the accreted matter.
A basic wind may not be the only physical process to lead to
mass ejection from the white dwarf. In the context of AGB and
super AGB star mass loss there may be an instability associated 
with local luminosity exceeding the Eddington limit that
leads to the expulsion of envelopes of small mass (Lau et al.
2012). It would be interesting to explore whether that sort
of instability operates in the context of binary transfer and,
if so, whether or not the magnetic field in the binary system 
affects it in any way. 


We have assumed here that the magnetic bottle funnels mass
from the M dwarf to the pole of the white dwarf, ignoring the
effect of an accretion disk that might, for instance, provide
a spin-up torque to the white dwarf. We can check the 
self-consistency of this assumption by estimating the 
location of the magnetosphere, $R_{mag}$, of the white dwarf based
on the dipole field of the white dwarf alone:
\begin{equation}
\label{magnetosphere}
R_{mag} \sim 5\times10^9 {\rm~cm} \left(\frac{B_{WD,6}^4R_{WD,9}^{12}}
      {m_{WD}\dot{m}_{WD,-7}^2}\right)^{1/7},
\end{equation}
where $\dot{m}_{WD,-7}$ is the accretion rate onto the white dwarf
in units of $10^{-7}$ \m\ y$^{-1}$. For this choice of fiducial 
parameters, the magnetosphere would be somewhat beyond the white 
dwarf surface, but this is a lower limit, given the added effect 
of the field of the nearby M dwarf. Thus, while this argument does 
not constitute proof, it is plausible that the field in the magnetic
bottle is strong enough to disrupt any accretion disk.    

The mass contained in the magnetic bottle, $M_{bot}$, can 
be estimated by equating the magnetic pressure with the 
gas pressure of the matter:
\begin{equation}
\frac{B^2}{8\pi} \sim 400~B_2^2~{\rm dyne~cm^{-2}} \sim nkT 
       \sim 4\times10^{11}\frac{m_{bot}T_4}{r_{bot,11}^3} 
         ~{\rm dyne~cm^{-2}},
\end{equation}
or
\begin{equation}
\label{mbottle}
m_{bot} \sim 10^{-9}\frac{B_2^2 r_{bot,11}^3}{T_4},
\end{equation}
where $m_{bot}$ is the mass in solar units, $B_2$
is the magnetic field in the bottle in units of 100 G,
$r_{bot,11}$ is a characteristic size of the bottle in
units of $10^{11}$cm and $T_4$ is a characteristic
temperature, assumed here for discussion purposes to be
comparable to the ionization temparature of hydrogen, 
of the matter in the bottle. 

The matter in this bottle could contribute to a circumstellar
environment in which variable Na D is observed (Patat et al.
2007; Simon et al. 2009; Blondin et al. 2009; Sternberg et al. 2011;
Dilday et al. 2012). This matter would be swept up by and would 
decelerate the SN ejecta. It thus might play some role in accounting 
for the high-velocity features (Wang et al. 2003) observed in 
about 80\% of SN~Ia (Marion, private communication,
2012). These features are characterized by lines of
the Ca II NIR triplet moving with a characteristic velocity 
of $\sim 22,000$ \kms, significantly above the typical photospheric
velocity. Gerardy et al. (2004) hypothesize that the high
velocity features represent circumstellar matter that is
accelerated by the SN ejecta. They point out that any
such matter must lie at a typical distance much less than
the radius of the photosphere near maximim light $\sim10^{15}$
cm, so that the collision does not contaminate the light
curve. They also estimate that to give the  typical velocity
of the high-velocity Ca II, the circumstellar medium must
have a mass of $\sim 0.02$\m. This is rather large compared to
the fiducial mass in Equation \ref{mbottle}, but such a
mass might be contained if the field were of order 1000 G,
as might characterize M dwarfs less than the transition
mass at MV4, and with a characteristic size of $r_{bot,11} \sim 50$.
It is not clear that a field of 100 G can be maintained at
$r_{bot,11} \sim 50$ even if the surface field of the M
dwarf is 1000 G. If the material were substantially cooler than 
10,000 K, these restrictions might be relaxed. 

It is also of interest to estimate the optical depth of 
the matter that might be contained in the magnetic bottle.
\begin{equation}
\label{depth}
\tau_{bot} \sim \kappa \rho \sim 50~\kappa\frac{B_2^2 r_{bot,11}}{T_4}.
\end{equation}
This suggests that the configuration might be optically thick,
and not at all resemble a naked WD/M pair just prior to any SN explosion.

\section{Mass Transfer and Loss}
\label{transfer}

Angular momentum loss from the WD/M system will be 
driven by gravitational radiation, by the losses associated
with a wind and magnetic braking from the magnetic bottle, by
electromagnetic energy from the rotating magnetic bottle, 
by magnetic braking from the M dwarf, and by direct mass loss
from the white dwarf. In what follows, the magnetic braking from 
the M dwarf itself and direct mass loss from the white dwarf will 
be ignored, for the reasons given above. The rotation of the combined
dipole field structure might generate electromagnetic radiation,
but the contribution to angular momentum loss will be trivially small.

The total angular momentum of the system is:
\begin{equation}
J_{TOT} = M_M a_M^2\Omega_{orb} + M_{WD}a_{WD}^2\Omega_{orb}
          + I_M\Omega_M + I_{WD}\Omega_{WD},
\end{equation}
where $a_M = (M_{WD}/M_{TOT}) a$ and $a_{WD} = (M_{M}/M_{TOT}) a$.
Invoking the tidal and magnetic torque locking described
above, $\Omega_M = \Omega_{WD} = \Omega_{orb}$. With $\Omega_{orb}
= (G M_{TOT}/a^3)^{1/2}$, the total angular momentum can thus be written:
\begin{equation}
J_{TOT} = M_M M_{WD} \left(\frac{G a}{M_{TOT}}\right)^{1/2} + 
        (I_M + I_{WD})\left(\frac{G M_{TOT}}{a^3}\right)^{1/2}.
\end{equation}
To account for the role of the moments of inertia, it is 
convenient to adopt mass-radius relations for the white dwarf
and the M dwarf. For the white dwarf we take 
\begin{equation}
\label{rwd}
\frac{R_{WD}}{R_{\rm\odot}} \sim 
       \left(\frac{M_{WD}}{M_{\rm\odot}}\right)^{-1/3},
\end{equation}
which is a decent approximation for lower mass white dwarfs, although
less so near the Chandrasekhar limit. For the M dwarf,
\begin{equation}
\label{rm}
\frac{R_{M}}{R_{\rm\odot}} \sim \frac{M_{M}}{M{\rm\odot}}.
\end{equation}
With these assumptions, the rate of change of the moments of inertia
can be written:
\begin{equation}
\dot{I}_{WD} = \frac{1}{3}I_{WD}\frac{\dot{M}_{WD}}{M_{WD}},
\end{equation}
and
\begin{equation}
\dot{I}_{M} = 3I_{M} \frac{\dot{M}_{M}}{M_{M}}.
\end{equation}
The rate of change in angular momentum of the orbit can thus be 
written as: 
\begin{eqnarray}
\label{jdot}
\dot{J}_{TOT} &= \left[M_M M_{WD} \left(\frac{G a}{M_{TOT}}\right)^{1/2}
  + \frac{1}{3}I_{WD}\left(\frac{G M_{TOT}}{a^3}\right)^{1/2}\right]
       \frac{\dot{M}_{WD}}{M_{WD}} \nonumber \\ 
              & + \left[M_M M_{WD} \left(\frac{G a}{M_{TOT}}\right)^{1/2} 
                + 3I_M\left(\frac{G M_{TOT}}{a^3}\right)^{1/2}\right]
                    \frac{\dot{M}_{M}}{M_{M}} \nonumber \\
              & + \left[-\frac{1}{2}M_M M_{WD}
                \left(\frac{G a}{M_{TOT}}\right)^{1/2} +
                \frac{1}{2}(I_M + I_{WD})
                \left(\frac{G M_{TOT}}{a^3}\right)^{1/2}\right]
                \frac{\dot{M}_{TOT}}{M_{TOT}} \nonumber \\
              & + \left[\frac{1}{2}
                 M_M M_{WD}\left(\frac{G a}{M_{TOT}}\right)^{1/2} -
                 \frac{3}{2}(I_M + I_{WD}
                 \left(\frac{G M_{TOT}}{a^3}\right)^{1/2}\right]
                 \frac{\dot{a}}{a}.   
\end{eqnarray}
The rate of change in the angular momentum must also be equal to
the total sinks of angular momentum, which we take to be:
\begin{equation}
\dot{J}_{TOT} = \dot{J}_{GR} + \dot{J}_{orb}, 
\end{equation}
where 
\begin{equation}
\dot{J}_{GR} = -4.5\times10^{34} {\rm~g~cm^2~s^{-2}} 
    \frac{m_M^2 m_{WD}^2 m_{TOT}^{1/2}}{a_{11}^{7/2}},
\end{equation}
and where $\dot{J}_{orb}$ represents the other modes of orbital 
angular momentum loss associated with mass loss from the system.  

\begin{figure}[htp]
\centering
\includegraphics[totalheight=0.5\textheight]{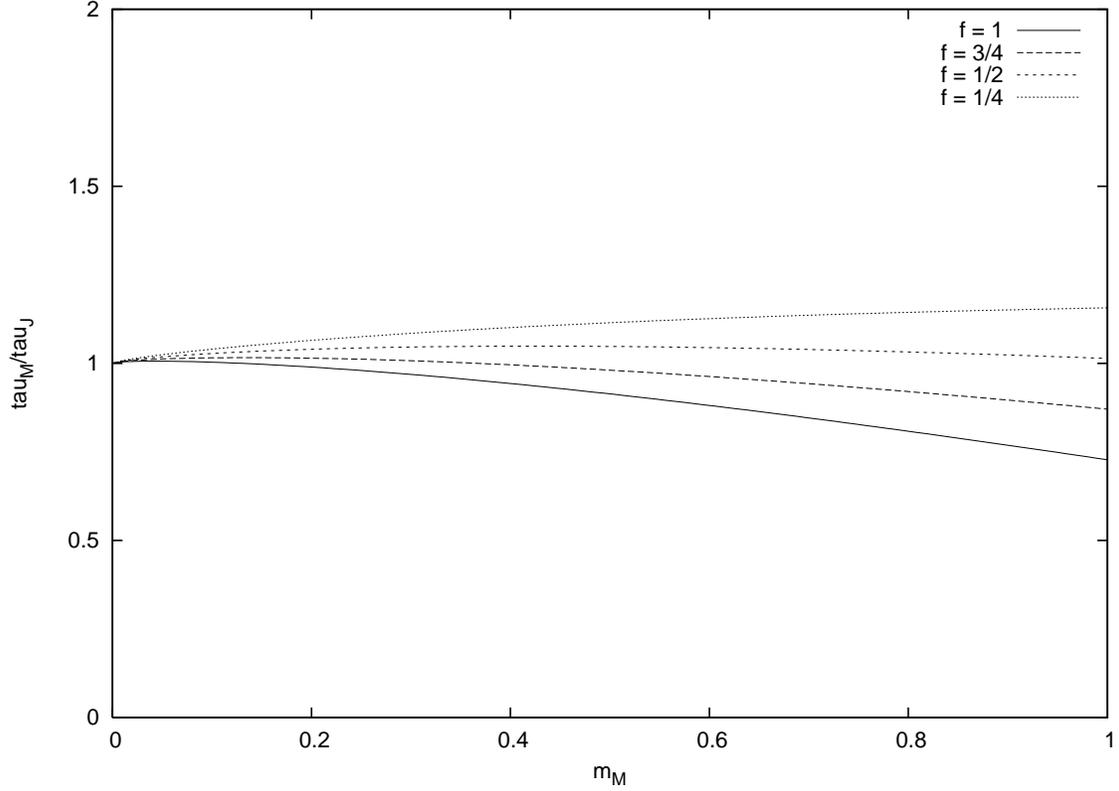}
\figcaption[jmratio.ps]
{The ratio, $\tau_M/\tau_{J_{orb}} \equiv 
(\dot{J}_{orb}/J_{orb})/(\dot{M}_{M}/M_{M})$
as a function of the mass of the M dwarf for various values of
the fraction of mass loss from the system, $f$, assuming
the mass of the white dwarf to be 1.4\m, the white dwarf and
the M star to be rotating synchronously, and for the M star
to be filling its Roche lobe. The moment of inertia of the
M dwarf is taken to be $I = kMR^2 = 0.4 M_{\odot} R_{\odot}^2 (m_M)^3$
\label{jmratio}}
\end{figure}

For illustration, let us assume that the M dwarf fills its
Roche lobe. We can then use equations \ref{Roche} and \ref{rm} 
to write the separation corresponding to filling the Roche lobe as:
\begin{equation}
\label{Roche_separation}
a_{RL,11} \sim 1.5~m_M^{2/3}m_{TOT}^{1/3},
\end{equation}
from which we can write:
\begin{equation}
\frac{\dot{a}}{a} = \frac{2}{3}\frac{\dot{M}_{M}}{M_{M}} +
      \frac{1}{3}\frac{\dot{M}_{TOT}}{M_{TOT}}. 
\end{equation}
Equation \ref{jdot} can then be written as:
\begin{eqnarray}
\label{jdotRL}
\dot{J}_{TOT} &= \left[M_M M_{WD} 
               \left(\frac{G a_{RL}}{M_{TOT}}\right)^{1/2}
  + \frac{1}{3}I_{WD}\left(\frac{G M_{TOT}}{a_{RL}^3}\right)^{1/2}\right]
   \frac{\dot{M}_{WD}}{M_{WD}}
\nonumber \\ 
              & + \left[\frac{4}{3}M_M M_{WD}
    \left(\frac{G a_{RL}}{M_{TOT}}\right)^{1/2} +
      (2I_M - I_{WD})\left(\frac{G M_{TOT}}{a_{RL}^3}\right)^{1/2}\right]
       \frac{\dot{M}_{M}}{M_{M}}
\nonumber \\
              & -\frac{1}{3}M_M M_{WD}
         \left(\frac{G a_{RL}}{M_{TOT}}\right)^{1/2}
          \frac{\dot{M}_{TOT}}{M_{TOT}}. 
\end{eqnarray}
We now assume that a fraction $f$ of the mass leaving the M dwarf 
arrives at the white dwarf and thus that a fraction $1 - f$ is 
ejected in some manner from the binary system. We can thus
write $\dot{M}_{WD} = -f\dot{M}_{M}$ and $\dot{M}_{TOT} = 
(1-f)\dot{M}_{M}$,
with the understanding that $\dot{M}_{M} < 0$. Neglecting the
moment of inertia of the white dwarf as being small compared
to that of the M dwarf, $I_{WD} << I_M$, neglecting $\dot{J}_{gr}$
so that $\dot{J}_{TOT} = \dot{J}_{orb}$, and  using
\begin{equation}
J_{orb} = M_M M_{WD}\left(\frac{G a_{RL}}{M_{TOT}}\right)^{1/2},
\end{equation}
Equation \ref{jdotRL} can be rewritten as:
\begin{equation}
\label{jdotRLloss}
\frac{\dot{J}_{orb}/J_{orb}}{\dot{M}_{M}/M_{M}} 
    \equiv \frac{\tau_M}{\tau_{J_{orb}}} = 
     -f\frac{M_M}{M_{WD}}\left(1 -\frac{1}{3}\frac{M_{WD}}{M_{TOT}}\right)
        + \frac{4}{3}-\frac{1}{3}\frac{M_{WD}}{M_{TOT}}
        + 2\frac{I_M M_{TOT}}{M_M M_{WD}a_{RL}^2}, 
\end{equation}    
where $\tau_M = M_{M}/\dot{M}_{M}$ is the characteristic timescale
for mass loss from the M star and $\tau_{J_{orb}} = 
J_{orb}/\dot{J}_{orb}$ is the characteristic timescale for loss 
of angular momentum from the system. The last term in Equation 
\ref{jdotRLloss} can be evaluated with the aid of Equation 
\ref{Roche_separation}. Evaluating the moment of inertia of the 
M dwarf as $I_M = kM_MR_M^2 \sim 0.4 M_{\odot} R_{\odot}^2 (m_M)^3$, 
that term can be written as $0.17 m_M^{2/3}m_{WD}^{-1}m_{TOT}^{1/3}$. 
Figure \ref{jmratio} gives the ratio, 
$(\dot{J}_{orb}/J_{orb})/(\dot{M}_{M}/M_{M})$ 
as a function of the mass of the M dwarf for $m_{WD} = 1.4$.

If we can specify the physics of the mass loss from the M dwarf,
then we can compute $\dot{J}_{TOT}$ and 
\begin{equation}    
\frac{\dot{a}}{a} = \left(\frac{2}{3} +
      \frac{1-f}{3}\frac{M_{M}}{M_{TOT}}\right)\frac{\dot{M}_{M}}{M_{M}},
\end{equation}    
with $f$ as a remaining free parameter. While the limit $f = 0$
makes physical sense in Equation \ref{jdotRLloss}, the limit
corresponding to all the mass leaving the M dwarf being expelled
from the system, the limit $f = 1$, is not valid here because 
we have assumed that the separation is such that Roche lobe contact 
is maintained. This demands that some net mass loss occur from the 
system. Note that we have here adopted a mass-radius relation for 
the M dwarf that corresponds to thermal equilibrium; that may not 
be the case, especially if the M dwarf is exposed to irradiation 
from the white dwarf. The key physical question is, what is 
$\dot{M}_{M}$?

\section{Irradiation and Mass Transfer}
\label{irradiation}

A major open issue of the model presented here, or any other
SD model, is whether the accretion onto the white dwarf
will actually grow the white dwarf to very near the Chandrasekhar
limit to trigger degenerate carbon ignition and explosion.
As mentioned in the introduction, one possibility is that
the systematics of accretion are such that little mixing
occurs from the inner white dwarf matter into the accreted
material. In this case, it might be possible to avoid
classical nova explosions that not only eject accreted matter,
but remove some of the white dwarf matter as well (Starrfield
et al. 2012). The presence of the magnetic field on the 
surface of the white dwarf might suppress such mixing.
The other option is to accrete sufficiently rapidly that
shell burning is in steady state and nova-like eruptions
are avoided (note that Starrfield et al. make the case that
steady burning is an artifact of insufficient resolution). 
We will also explore that option by investigating the
liklihood that a WD/M system may provide the mass transfer 
rate to yield non-degenerate H, He shell burning, beat the 
nova limit, and grow the white dwarf to central carbon ignition. 

One option to promote steady burning is to consider that 
the mass loss may be channeled by the magnetic bottle connecting 
the two stars. In this situation, the mass may thus land on 
a concentrated polar area of the white dwarf, enhancing the 
effective local rate of accretion compared to spherical 
accretion. The polar area is estimated to be $f_{pole} \sim 1\%$ 
of the white dwarf surface for typical polars (O'Donoghue et al. 
2006), so the effective enhancement of the local accretion rate 
per unit area might be of order 100. In addition, the magnetic 
field may inhibit shear mixing and convection in a manner such as to 
weaken any explosion (Livo, Shankar \& Truran 1988). 

Another interesting possibility is that X-ray illumination 
from the pole cap of the white dwarf pointed at the M dwarf might 
drive self-sustained mass transfer. Schaefer \& Collazi (2010)
identify a class of novae comprising $\sim 14\%$ of all novae, 
named after the prototype, V1500 Cyg, that are characterized 
by the post-eruption magnitude being much brighter than before 
eruption, short orbital periods, long-lasting emission from the 
white dwarf following the eruption, a highly magnetized white dwarf, 
and a secular decline in the post-eruption luminosity. Schaefer
et al. (2011) conclude that the recurrent nova T Pyx, one of
the V1500 Cyg systems, is likely to be a polar, with the source
of the optical luminosity arising from non-thermal sources, 
rather than an accretion disk. Following van Teesling \& King (1998), 
King \& van Teesling (1998) and Knigge et al. (2000), Schaefer 
\& Collazi advocate that bloating of the companion and a 
self-excited wind, driven by accretion and burning at the 
magnetic pole can account for the properties of the V1550 Cyg systems.  
Binary neutron star analogs for this process are known (Fruchter
et al. 1988; Romani 2012).

Here we outline the systematics of such a self-excited wind.
Models with steady shell burning on the surfaces of white dwarfs 
produce a luminosity of about $L = 10^{37}L_{37}~$erg s$^{-1}$. 
If this energy is radiated from the accreting pole of the white 
dwarf, it may radiate into a restricted solid angle $\Delta\Omega 
< 4\pi$. The M star would then intercept a fraction of this luminosity
corresponding to $\pi R_M^2/\Delta\Omega a^2$. Using Equation \ref{Roche},
the luminosity incident on the M dwarf can then be written:
\begin{equation}
L_{in} \sim  5.3\times10^{35}~{\rm erg~s^{-1}}\frac{4\pi}{\Delta\Omega}
     L_{37}\left(\frac{M_M}{M_{TOT}}\right)^{2/3}
      \left(\frac{a_{RL}}{a}\right)^2,
\end{equation}
where the final factor allows for the possibility that the
M dwarf does not fill its Roche lobe. The mechanical power 
involved in producing a wind from the surface of the M dwarf is:
\begin{equation}
L_{wind} \sim \frac{GM_M\dot{M}_M}{R_M} \sim 1.9\times10^{15} 
   ~{\rm erg~s^{-1}}\dot{M}_M,
\end{equation}
where we have taken $M_M/R_M = M_\odot/R_\odot$. With
$L_{wind} = \epsilon L_{in}$, where $\epsilon$ is the 
efficiency to turn incident irradiation into wind, we have:
\begin{equation}
\label{radmdot}
\dot{M}_M \sim 2.8\times10^{20}~{\rm g~s^{-1}}
    \epsilon L_{37} \frac{4\pi}{\Delta\Omega}
        \left(\frac{M_M}{M_{TOT}}\right)^{2/3}
           \left(\frac{a_{RL}}{a}\right)^2.
\end{equation}
The source of the luminosity might be the accretion or that
from burning shells. For accretion, recalling that the
mass accretion rate onto the white dwarf is a fraction, $f$, 
of that leaving the M star, we have:
\begin{equation}
L_{acc} = \frac{GM_{WD}f\dot{M}_M}{R_{WD}} \sim 1.3\times10^{37}
   ~{\rm erg~s^{-1}} \frac{m_{WD}f\dot{M}_{M}}{R_{WD,9}}.
\end{equation}
Invoking the mass loss from the M dwarf from Equation \ref{radmdot},
we have:
\begin{equation}
\frac{a}{a_{RL}} \sim 1.9 \left(\frac{m_{WD}}{R_{WD}}\right)^{1/2}
      \left(\frac{M_M}{M_{TOT}}\right)^{1/3}
        \left(\frac{4\pi f \epsilon}{\Delta\Omega}\right)^{1/2}.
\end{equation}
If the incident luminosity is from a hydrogen-burning shell,
\begin{equation}
L_{burn} = 0.007 X_H \dot{M}_{WD} c^2 = 5.6\times10^{18}~{\rm erg~s^{-1}}
     X_H \dot{M}_{WD},
\end{equation}
where $X_H$ is the mass fraction of hydrogen in the accreted
matter (ignoring any luminosity from helium shell burning), 
a similar argument gives:
\begin{equation}
\frac{a}{a_{RL}} \sim 13\left(\frac{M_M}{M_{TOT}}\right)^{1/3}
        \left(\frac{4\pi f \epsilon X_H}{\Delta\Omega}
           \right)^{1/2}.
\end{equation}

\begin{figure}[htp]
\centering
\includegraphics[totalheight=0.5\textheight]{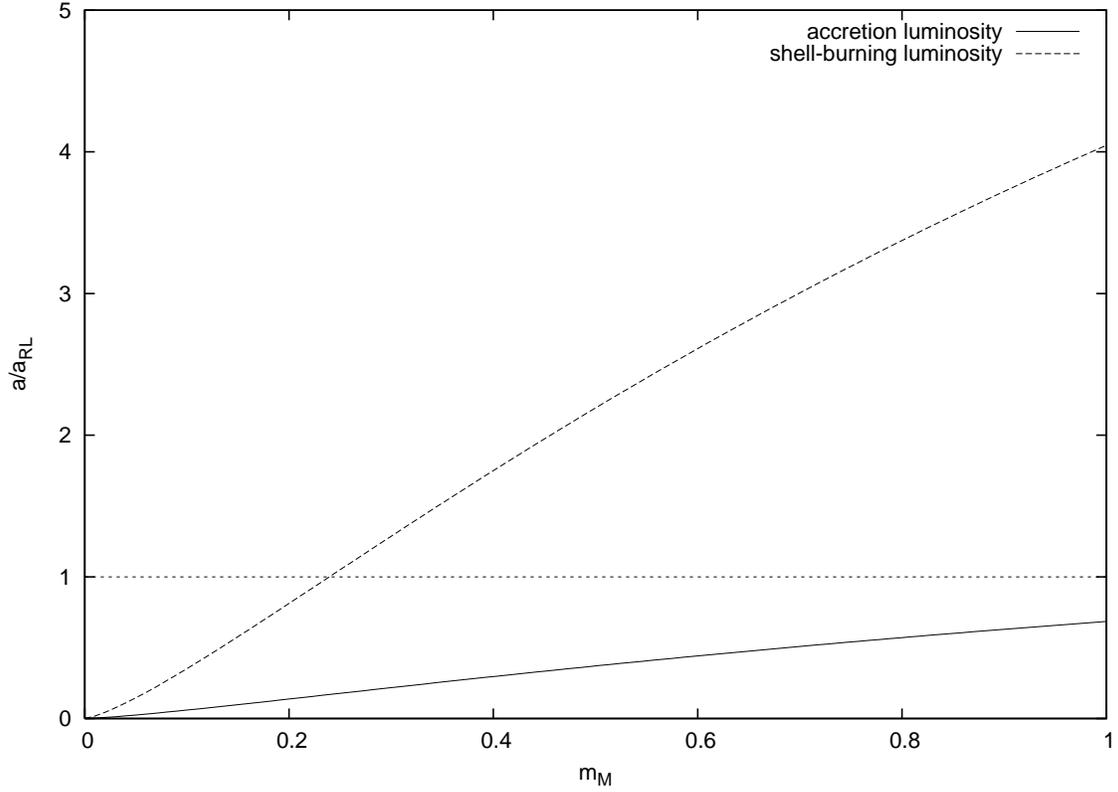}
\figcaption[irrad.ps]
{The ratio, $a/a_{RL}$ as a function of the mass of the M dwarf 
by which the separation can exceed that corresponding to filling
the Roche lobe and maintain a self-consistent, self-irradiated 
condition due to accretion luminosity (solid line) or hydrogen
shell burning (dashed line). The curves are constructed assuming 
$M_{WD} =1.4$\m, $R_{WD,9} = 1$, and efficiency factor (see text) 
equal unity. To be physical, $a/a_{RL}$ must exceed unity, as 
indicated by the horizontal line. This condition is readily
met by shell-burning luminosity, but for accretion luminosity
alone to be self-consistent either the radius of the white dwarf
would need to be less than $10^9$~cm and/or the efficiency factor
would need to be greater than one, a possible condition for
irradiation into a narrow solid angle from the pole cap of 
the white dwarf.  
\label{irrad}}
\end{figure}

Figure \ref{irrad} gives the ratio, $a/a_{RL}$, as a function of 
the mass of the M dwarf that satisfies the condition for 
self-consistent, self-irradiation for the cases where the
luminosity is provided by accretion or by hydrogen shell burning.
The curves are presented in Figure \ref{irrad} assuming $M_{WD} =
1.4$\m\ and $R_{WD,9} = 1$. One can define efficiency factors
$\Phi_{acc} = \left(\frac{4\pi f \epsilon}{\Delta\Omega}\right)^{1/2}$  
and $\Phi_{burn} = \left(\frac{4\pi f \epsilon X_H}{\Delta\Omega} 
\right)^{1/2}$, both of which are taken to be unity in Figure \ref{irrad}.
To be physical, the condition for self-irradiation, $a/a_{RL}$, 
must exceed unity. This condition is readily met by shell-burning 
luminosity down to rather low values of the mass of the M dwarf. 
For the chosen parameters, accretion luminosity alone would not be 
self-consistent. The accretion model would be more consistent if the 
radius of the white dwarf were less than $10^9$cm, a likely condition
as the white dwarf approaches the Chandrasekhar limit. Another
possibility would be for the efficiency factor, $\Phi_{acc}$, to 
exceed unity.  This might also be possible if the solid angle into 
which the white dwarf pole cap radiated were sufficiently small to 
overcome the mass loss from the system ($f < 1$) and the efficiency 
to convert incident radiation into mass loss. 

This analysis implies that a self-excited wind based on 
accretion power alone might be self-consistent if the M dwarf 
nearly fills its Roche lobe. If the source of the luminosity 
is a steady hydrogen-burning shell, then the M dwarf could
be more widely separated. In that circumstance, one can 
envisage the possibility that the M dwarf is brought to the
condition that it fills its Roche lobe, starting the
mass transfer and burning process. In a transient phase, the
M dwarf might retreat as it lost mass, until a quasi-steady
state, shell-burning self-excited wind were established.
A typical transfer rate might be $\sim 10^{20}$ g s$^{-1}$
or $\sim 10^{-6}$\m\ y$^{-1}$. Such a transfer rate might well
sustain shell burning, especially if the burning were confined
to the small area of the magnetic pole of the white dwarf.  
Even if such a self-excited wind can be established, the details 
clearly depend on the the solid angle of X-radiation from the white 
dwarf pole cap, the mass loss fraction, f, from the system and 
the efficiency of conversion of incident irradiation to mass loss 
from the M dwarf. It is also possible that the transfer stream is 
self-shielded, yielding a time-dependent flow. For a more detailed, 
general analysis of self-excited winds, see King \& van Teesling (1998) 
and references therein.  Published analyses of self-excited winds 
do not seem to account for the possibility that the companion star, 
as well as the white dwarf, is magnetized.

Earlier estimates suggested that the matter in the magnetic 
bottle might be optically thick. If that were to be the case, 
an estimate can be made of the effective temperature:
\begin{equation}
T_{eff} \sim 3.4\times10^4 \frac{L_{37}^{1/4}}{r_{bot,11}^{1/2}},
\end{equation}
were $r_{bot,11}$ is the characteristic size of the
photosphere of the bottle. The effective temperature might thus
be on the order of the ionization temperature of hydrogen for 
the level of luminosity discussed here and the bottle comparable
in size to the orbital separation. Adopting the
expression for the optical depth from Equation \ref{depth} the
effective temperature can be written as:
\begin{equation}
T_{eff,4} \sim 24 B_2 L_{37}^{1/4}\left(
      \frac{\kappa}{\tau_{bot}T_4}\right)^{1/2}.
\end{equation}
It is thus plausible to have $T_{eff} < T \sim 10^4$K for 
$\tau_{bot} \ga 10$ and $\kappa < 1$ g~cm$^{-2}$.
The effective temperature of the bottle can also be
expressed as a function of the mass in the bottle using
Equation \ref{mbottle} as:
\begin{equation}
T_{eff,4} \sim 0.1 B_2^{1/3} L_{37}^{1/4}m_{bot}^{-1/6}T_4^{-1/6}.
\end{equation}

Schaefer \& Collazzi (see also Schaefer et al. 2010 and 
Schaefer, Pagnotta \& Shara 2010) argue that the V1500 Cyg 
systems cannot become SN~Ia. Some are neon novae, certainly 
counterindicative. Some have total mass less than the
Chandrasekhar mass. These factors whittle down
the eligible population, but do not preclude that there
remains a finite fraction of such systems that do contain
C/O white dwarfs and a total mass exceeding the Chandrasekhar
mass. Another argument is that the post-outburst luminosity
in the V1500 Cyg systems fades on timescales of decades to 
centuries, suggesting that the systems are heading for hibernation 
(Shara et al. 1986). While it is hardly a rigorous counter 
argument, we point out while we have invoked magnetic locking 
of the poles for simplicity, the expectation is that the 
magnetic poles of the white dwarf and M dwarf will oscillate 
about co-alignment with a timescale that is also comparable 
to decades to centuries, Equations \ref{time,wd} and \ref{time,m}. 
The secular evolution might be associated with that phenomenon, 
rather than an evolution toward hibernation. If that were the 
case, one would also expect to see systems with secularly rising 
luminosity. The statistics of catching a system in that phase are 
not clear. If the SN~Ia progenitor systems do undergo a secular 
oscillating accretion onto the white dwarf and illumination of 
the M dwarf, the issue of the net white dwarf accretion rate 
becomes more complex.

While the V1500 Cyg systems have certain generic properties
that are reminiscent of the SN~Ia progenitor systems explored 
here, magnetic white dwarfs, M dwarf companions, absent or 
truncated accretion disks, and apparent self-irradiated 
mass transfer, it is not clear that they represent the
progenitor systems being sought. Rather they might be close
cousins. It is possible that if similar systems do become
locked into a truly quasi-stable self-irradiated transfer
state that they appear differently. The obvious possibility is
that the matter in the magnetic bottle becomes optically
thick, thus ``cloaking" an otherwise similar system. 
 
\section{Beating the Companion Luminosity Limits}
\label{post}

One of the original goals of this investigation was to
explore the possibility of beating the observed companion 
luminosity limits. Here, we re-visit some of
the factors involved. The case has been made that the
luminosity of most or all isolated M dwarfs is less than 
the Schaefer/Pagnotta/Schmidt limit on companion stars
that survive the explosion. It is worth noting that these 
limits apply after all the proposed processes of mass transfer 
and loss and after the explosion. It is reasonable to suppose 
that the pertinent systems start, after the relevant 
common-envelope phase, with higher mass and hence more chance 
for the total mass of the system to exceed the Chandrasekhar mass.  
The companion only needs to be dimmer than the relevant
limits by the time of the explosion. Strictly speaking, the 
companion only needs to be dimmer than the relevant limits 
several hundred years after the explosion for the cases of 
SNR 0509-67.5, Tycho's supernova and SN~1006. There are two 
issues involved here, the mass stripped in the explosion and 
the luminosity of the M dwarf after the impact.

The stripping of mass from a companion star by a supernova 
has been considered by Livne, Tuchman \& Wheeler (1992), 
Marietta, Burrows \& Fryxell (2000), Pakmor et al. (2008),
Kasen (2010) and Pan, Ricker \& Taam (2012), among others. 
When considering main sequence companions,
these works have all taken the companion to be $\sim 1$\m,
following standard SD paradigms. Such stars have small
convective envelopes with substantial radiative cores. 
The density structure will be fairly centrally condensed,
similar to a polytrope of index n = 3. M dwarfs, by 
contrast, will be primarily, even totally convective. They
will have a density structure that is less centrally
condensed and better described by a polytrope of n = 3/2.
This ``puffier" structure (even neglecting departures from
radiative equilibrium from mass transfer and irradiation)
may alter the fraction of mass stripped from the M star.  
From Equation \ref{Roche}, the fraction of the ejecta
intercepted by an M dwarf that is filling its Roche lobe, $f_{ej}$, is:
\begin{equation}
f_{ej} = \frac{1}{4}\left(\frac{R_M}{a_{RL}}\right)^2
       \sim 5.3\times10^{-2}\left(\frac{M_M}{M_{TOT}}\right)^{2/3}. 
\end{equation}
Thus an M dwarf with a mass of several tenths of \m\ will
intercept about $10^{49}$ ergs of energy. Reality is more
complex, however, with the shock parting around the companion,
so only some of the matter is stripped, some ablated (Wheeler,
Lecar \& McKee 1975; Pan, Ricker \& Taam 2012), and
only some of the incident energy goes to heating the star.
For perspective the binding energy of an M dwarf, $E_M$, is:
\begin{equation}
E_M \sim \frac{GM_M^2}{R_M} \sim 2.5\times10^{48}m_M~erg.
\end{equation}
There is sufficient energy intercepted to unbind a small mass 
M dwarf so no companion survives, but this needs to be investigated
quantitatively. 

If a portion of the companion survives, it is likely that it 
will be substantially heated and hence of higher luminosity for 
a Kelvin-Helmholz cooling time. Taking the radius to be 
proportional to the mass and the luminosity to be proportional
to $M^4$, the Kelvin-Helmholz time is $\tau_{KH} \sim GM_M^2/R_ML_M
\sim 3\times10^7 m_M^{-3}$ y. This is nominally a quite long time
compared to the timescale on which the post-explosion observations 
of SN~Ia companions have been made. This issue may require
deeper study since this timescale is basically estimated for
a star in thermal equilibrium. A possibly important consideration
is that M dwarfs are highly or fully-convective. The convection
can rapidly move flux within the star, on something like the
dynamical timescale. The star cannot, of course, convect energy
into space, so the luminosity is set by the outer boundary conditions
and hence sensitive to the opacity. If such a star is shocked
by a supernova and heated, but remains convective, the opacity in
the outer layers could change substantially, perhaps declining
from rather large values characteristic of molecular opacities to
something like electon scattering. In such a situation, the
cooling time in the immediate aftermath of a supernova shock
might be less than the simple estimate above would indicate.


\section{Conclusions and Discussion}
\label{concl}

The discussion here scarcely constitutes proof that SN~Ia
arise in the explosion of white dwarfs with M dwarf companions,
but it does illustrate that an interesting case can be made
for further investigation. Because much of the mass of the 
M dwarf may be consumed in the hypothesized process, the 
proposed model could be described as a ``single-degenerate
coalescence model." The salient points are:
\begin{itemize}
\renewcommand{\labelitemiv}{$\ast$}
\item M dwarfs are of sufficiently low luminosity to evade the
limits on left-over companions of SN~Ia progenitors in young
supernova remnats.
\item white dwarfs and M dwarfs are the most common stellar objects 
in the Galaxy. Their binary combination is also quite common. 
Coalescence of white dwarfs with M dwarf companions could plausibly 
occur at a rate comparable to the observed rate of SN~Ia.
\item $\sim 20\%$ of white dwarfs have mass in excess of 0.8\m. If 
such white dwarfs consume nearly all the mass of an M dwarf, they
can, in principle, reach the Chandrasekar mass.
\item M dwarfs do not evolve in the absence of binary mass transfer.
The delay time distribution of coalescing WD/M pairs could be dominated
by gravitational radiation and resemble the observed distribution
for SN~Ia and that hypothesized for the double degenerate models of SN~Ia.
\item M dwarfs are magnetic and some fraction of their white 
dwarf companions will also be. Mass loss by flares is modest,
but gives a hard floor to the mass loss by M dwarfs.
\item M dwarfs of smaller mass, those later than spectral type MV4,
tend to have substantially larger magnetic fields. This transition
may play a role as mass is lost in a binary system.
\item The stable equilibrium of two magnetic stars is for the 
opposite poles to align with one another in the orbital plane,
perhaps oscillating around this equilibrium. The timescale to
reach this equilibrium is short. 
\item Tidal and magnetic torques will lock both the white dwarf
and the M dwarf into synchronism with the orbit. The white
dwarf is expected to be {\it slowly} rotating.
\item The combined magnetic fields of the two co-aligned stars
will produce a ``magnetic bottle" that will affect the transfer
of matter between them.
\item This magnetic bottle may prevent the formation of a wind
from the white dwarf hypothesized in some rapidly-accreting models 
of SN~Ia.
\item The mass in the magnetic bottle might play some role
in forming the circumstellar matter seen in SN~Ia with variable
NaD and in the formation of the high-velocity Ca II feature
at $\sim 22,000$\kms.
\item The mass in the magnetic bottle may be optically thick so
the system just prior to explosion may not resemble a ``naked"
cataclysmic variable.
\item The system could lose some fraction of the mass transferred
from the M dwarf and associated angular momentum at a rate
sufficient to drive the M dwarf into contact and to maintain
that situation.
\item Close binaries of the sort investigated here may be 
susceptible to self-excited mass loss from the M dwarf with
the luminosity deriving either from the accretion luminosity
from the white dwarf or the luminosity of nuclear shell
burning on its surface.
\item The self-excited mass transfer might be sufficient to
maintain shell burning on the white dwarf. This is especially 
true if the accretion column is magnetically focused into a 
small polar area on the white dwarf (that is selectively aimed 
back at the M dwarf as outlined above) and if turbulent mixing 
of that matter both laterally and with the inner white dwarf 
matter is inhibited by the magnetic field of the white dwarf.  
\item The class of V1500 Cyg variables, prominantly including
T Pyx, have many of the features invoked in the models
presented here.
\item The mass of the companion star does not have to be
especially small initially to beat the observed limits on
the surviving companions of SN~Ia. It only needs to be
sufficiently dim by the end of the mass transfer process.
\item M dwarfs are likely to be nearly fully or completely
convective, so have a less centrally concentrated structure 
than usually considered for the companions in the SD scenario. 
They may thus be more susceptible to mass stripping and
disruption in the explosion than normally considered.  
\ Because they are nearly or fully convective, M dwarfs
may quickly radiate any heat gained by the supernova
impact and thus may have settled back into dim
thermal equilibrium hundreds or thousands of years later
when the observations constraining companions have been made. 
\end{itemize}

Having made these points of interest, there are still 
many topics that require more careful study.
\begin{itemize}
\renewcommand{\labelitemiv}{$\ast$}
\item What is the distribution of WD/M pairs, their
masses, separations and magnetic fields? This area
is ripe for more observational study, but also 
requires investigation with appropriate detailed
binary synthesis codes to characterize and constrain
the nature of the associated common envelope evolution
and the resulting distributions.
\item What is the nature of the combined magnetic fields
of a white dwarf and an M dwarf companion, how does it
affect mass transfer and loss from the system?
\item How do the combined magnetic fields and that on the
white dwarf affect the accretion and burning of the matter
on the surface of the white dwarf?
\item Can radiation-induced self-excited mass transfer be sustained?
\item How does the mass of the M dwarf evolve during 
mass transfer? Does the field increase when the mass
is peeled down to the M4V transition mass? Does the
magnetic field affect the mass at which this transition
occurs?
\item Does tidal locking enhance the M dwarf dynamo?
\item What would an enshrouded, magnetically-bottled system 
look like prior to explosion?
\item What is the configuration of the M dwarf companion
at the time of explosion in terms of mass, magnetic field,
radius?
\item What is the effect of the explosion on the M dwarf,
and is any expelled hydrogen within observational limits?
\end{itemize}

There are many other complications to ponder in terms of
the white dwarfs, the M dwarfs and their pairing.

About 10\% of all white dwarfs are magnetic, with fields in excess of 
$\sim10^6$ G, while about 25\% of those in cataclysmic variables are 
magnetic (Wickramasinghe \& Ferrario 2000; Wickramasinghe \& Ferrario 2005).
The magnetic fields range from $\sim3\times10^4-10^9$ G with a 
distribution peaking at $1.6\times10^7$ G in the greater sample 
and from $\sim10^7-3\times10^8$ G in CVs. The magnetic white dwarfs 
tend to have a higher mass than the nonmagnetic white dwarfs, with
a mean mass of $\sim 0.93$\m (Wickramasinghe \& Ferrario 2005). 
Ferrario \& Wickramasinghe (2005) concluded that magnetic white dwarfs 
can be divided into three groups: those that are strongly magnetized 
slow rotators arising in single-star evolution, a group of strongly-magnetized 
fast rotators arising in mergers, and a group of modest rotators 
that have a mixed origin in single star and CV-like systems.
The surface field distributions tend not to be dipolar and can be 
modeled by dipoles that are offset by 10\%-30\% of the stellar 
radius along the dipole axis (Wickramasinghe \& Ferrario 2000). 

Liebert et al. (2005) examined the SDSS sample and the sample
from common proper motion systems of white dwarfs with 
evidence of magnetic field and detached M dwarf companions. 
They found no example where
a magnetic white dwarf was paired with a detached M dwarf, whereas
about 25\% of accreting systems have a magnetic white dwarf. 
They speculate that the presence of the companion and the relatively 
large mass and small radius of the magnetic white dwarf relative to 
nonmagnetic degenerate dwarfs yields a selection effect against 
the discovery of the magnetic white dwarfs in detached systems. 
The question of whether or not the mass and field distributions of 
the magnetic white dwarfs in interacting binaries are similar to 
those isolated magnetic white dwarfs remains to be determined.

As pointed out by Stassun et al. (2010) the relationships for M 
dwarfs between rotation, field strength, interior structure, and 
surface field topology are surely complex, non-linear, and dependent 
on the age of the star, even neglecting the effects of binary tidal 
locking and mass loss. Donati et al. (2008) and Reiners \& Basri 
(2008) have shown that the fraction of the magnetic field in large 
scale components is larger in late-type than in early type M dwarfs. 
The field structure changes from globally weak but highly structured 
multipolar and non-axisymmetric fields in early spectral types, to 
globally strong and ordered dipole fields in the later spectral 
types.  Browning (2008) produced models of fully convective stars
that generated large scale, long-lived magnetic fields. The affect 
of large fields on the structure and evolution of M dwarfs has been 
studied by Mullan \& McDonald (2001) and McDonald \& Mullan (2009).

Progress on the demographics of WD/M pairs and related
systems is promised by current wide field photometric
surveys such as PTF and PanSTAARS. Birkby et al. (2012)
studied detached M dwarf eclipsing binaries. The orbital 
periods fell in the range $1.5 < P< 4.9$ days. The 
masses ranged from 0.35-0.50\m\ with radii between 0.38-0.50 $R_\odot$. 
Birkby et al. argue that the tidally locking should give rise to 
high rotational velocities and high levels of magnetic activity that 
inflate the stellar radii and increase star spots (and flare mass loss?). 
They found that a radius inflation of order 10\% was manifested
even in non-synchronous M dwarfs of low activity. 

There is clearly still much to learn about the nature of M dwarfs 
in general and those in binary systems in particular. Wu
\& Wickramasinghe (1993) investigated equilibrium configurations 
and magnetic interactions in WD/M pairs for which the M dwarf 
was assumed to have an intrinsic centered dipole field, and 
the white dwarf an offset dipole or a centered dipole plus a 
quadrupole field. The interaction of two magnetic stars has 
been considered by Piro (2012) in the context of binary neutron stars.

There are indications that the white dwarf/M dwarf systems 
will not necessarily involve standard Roche lobe overflow mass 
transfer and intriguing issues to explore in the context
of the progenitor systems of SN~Ia.

\acknowledgments

The author is grateful to Brad Schaefer, Ashley Pagnotta, 
Rosanne DiStefano, Rob Robinson, Don Winget, Suzanne Hawley,
Keivan Stassun, Alicia Aarnio, John Tonry, and Brian Schmidt 
for constructive discussions and to David Branch and Alexei
Poludnenko for comments on the manuscript. This research is 
supported in part by NSF AST-1109801. Some work on this
paper was done in the hospitable environment of the Aspen 
Center for Physics that is supported by NSF Grant PHY-1066293.


{}


\newpage


\end{document}